\documentclass[english,prd,twocolumn,superscriptaddress,nofootinbib,preprintnumbers]{revtex4}

\usepackage[latin1]{inputenc}
\usepackage{graphicx}
\usepackage{amssymb}
\usepackage{amsfonts}
\usepackage{dcolumn}
\input{epsf}

\newcommand{\be}{\begin{equation}}
\newcommand{\ee}{\end{equation}}
\newcommand{\beq}{\begin{eqnarray}}
\newcommand{\eeq}{\end{eqnarray}}

\newcommand{\rd}{{\rm d}}

\makeatother

\usepackage{babel}
\makeatother

\begin{document}

\title{Background cosmological dynamics in  $f(R)$ gravity and observational constraints}
\author{Amna Ali}
\affiliation{Centre of Theoretical Physics, Jamia Millia Islamia, New Delhi-110025, India}

\author{Radouane Gannouji}
\affiliation{IUCAA, Post Bag 4, Ganeshkhind, Pune 411 007, India}

\author{M. Sami}
\affiliation{Centre of Theoretical Physics, Jamia Millia Islamia, New Delhi-110025, India}

\author{Anjan A. Sen}
\affiliation{Centre of Theoretical Physics, Jamia Millia Islamia, New Delhi-110025, India}

\begin{abstract}

In this paper, we  carry out a study of  viable cosmological models
in $f(R)$-gravity  at the background level. We use observable
parameters like $\Omega$ and $\gamma$ to form autonomous system of
equations and show that the models under consideration exhibit two
different regimes in their time evolution, namely, a phantom phase
followed by a quintessence like behavior. We employ statefinder
parameters to emphasize a characteristic discriminative signature of
these models.

\end{abstract}

\date{\today}

\maketitle

\section{Introduction}
One of the most challenging problems of modern  cosmology today is
associated with the attempts to understanding the late time
acceleration of Universe which is supported by cosmological
observations of complimentary nature such as Supernovae Ia
\cite{Perlmutter_Riess}, Cosmic Microwave Background anisotropies
\cite{Spergel:2006hy}, Large Scale Structure formation
\cite{Seljak:2004xh}, baryon oscillations \cite{Eisenstein:2005su}
and weak lensing \cite{Jain:2003tba} observations. Many theoretical
approaches have been employed to explain the phenomenon of late time
cosmic acceleration. The standard lore  assumes the presence of an
exotic fluid known as dark energy. The simplest dark energy model
based upon cosmological constant dubbed $\Lambda$CDM model suffers
from extreme fine-tuning and coincidence problems \cite{rev}. Scalar fields
minimally coupled to gravity, called quintessence, with generic
features might allow to alleviate these problems \cite{scalar}.
Many other possibilities have
been proposed, including
a scalar field with a non-standard kinetic term
($k$-essence)
\cite{Arm1,Garriga,Chiba1,Arm2,Arm3,Chiba2,Chimento1,Chimento2,Scherrerk},
or simply an arbitrary barotropic fluid with a pre-determined form
for $p(\rho)$, such as the Chaplygin gas and its various
generalizations \cite{Kamenshchik,Bilic,Bento,Dev,Gorini,Bean,Mul,Sen}.

As an alternative to dark energy, the large scale modifications of
gravity could account for the current acceleration of universe. We
know that gravity is modified at short distance and there is no
guarantee that it would not suffer any correction at large scales
where it is never verified directly. Large scale modifications might
arise from extra dimensional effects or can be inspired by
fundamental theories. They can also be motivated by phenomenological
considerations such as $f(R)$ theories of gravity (see
\cite{Sotiriou:2008rp} for a recent review). However, any large
scale modification of gravity should reconcile with local physics
constraints and should have potential of being distinguished from
cosmological constant.

Most of the f(R) gravity models proposed in the literature either
ruled out by cosmological constraints imposed by the
history\cite{Amendola:2006we} or fail to meet the local gravity
constraints \cite{LGC}. The viable f(R) models can be distinguished
from the $\Lambda$CDM by studying the the evolution of the growth of
matter density perturbations
\cite{Tsujikawa:2009ku,Gannouji:2008wt,Gannouji:2008jr}.

  In this paper, we explore the possibility of discriminating
the $\Lambda$CDM  from viable models of $f(R)$ at the background
level (see also Ref.\cite{ON} on the related theme).

\section{$f(R)$ cosmology}
In what follows, it would be convenient to us to write f(R) gravity
action in the form
 \be \label{eq:action} \mathcal{S}=\frac{1}{16\pi G}\int{\rm d}^4
x\sqrt{-g}\left[R+\epsilon(R)\right]+\mathcal{S}_m(g_{\mu
\nu},\Psi_m) \ee

where $G$ is the bare gravitational constant, $\epsilon(R)$ is a function of the curvature scalar $R$
 only and  $\mathcal{S}_m$ is a functional of some matter fields $\Psi_m$
 and  metric $g_{\mu \nu}$.\\
In case of flat homogenous and isotropic universe
 \be
\mathcal{\rd}s^2=-\mathcal{\rd}t^2+a(t)^2\left(\mathcal{\rd}r^2+r^2\mathcal{\rd}\Omega^2\right)
\ee

the action (\ref{eq:action}) gives rise to the following evolution
equations (we use the unit, $8\pi G=1$)

\beq
\label{eq:fried1}
3H^2&=&\rho_m+\frac{R\epsilon_{,R}-\epsilon}{2}-3H\dot{\epsilon}_{,R}-3H^2\epsilon_{,R}\\
\label{eq:fried2}
-2\dot{H}&=&\rho_m+\ddot{\epsilon}_{,R}-H\dot{\epsilon}_{,R}+2\dot{H}\epsilon_{,R}\\
\label{eq:conserv}
\dot{\rho}_m&+&3H\rho_m=0
\eeq

The equations (\ref{eq:fried1},\ref{eq:fried2}) can be rewritten with the definition of the density of dark energy

\beq
3H^2&=&\rho_m+\rho_{DE}\\
-2\dot{H}&=&\rho_m+\rho_{DE}(1+w_{DE}) \eeq where
$w_{DE}=P_{DE}/\rho_{DE}$.

\begin{figure*} \centering
\begin{center}
$\begin{array}{c@{\hspace{0.4in}}c}
\multicolumn{1}{l}{\mbox{}} &
        \multicolumn{1}{l}{\mbox{}} \\ [0.0cm]
\epsfxsize=3.2in
\epsffile{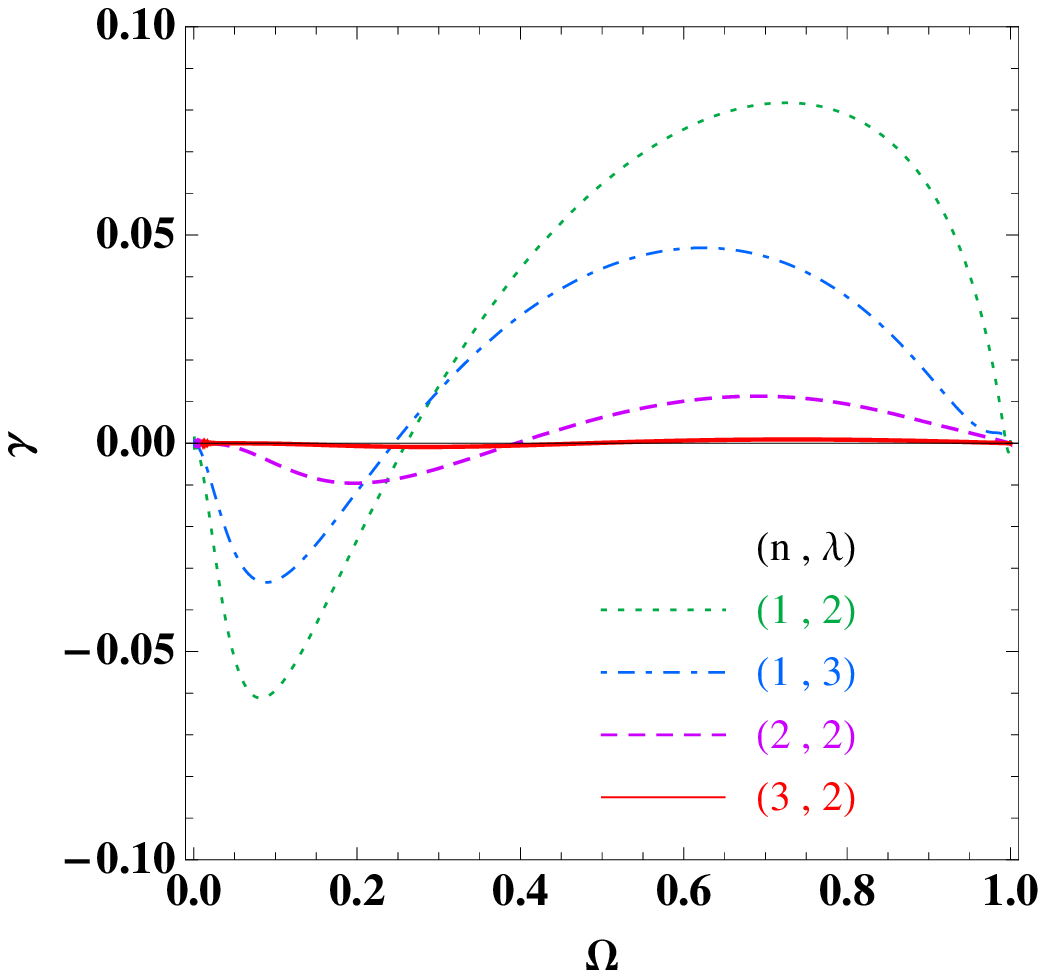} &
        \epsfxsize=3.2in
        \epsffile{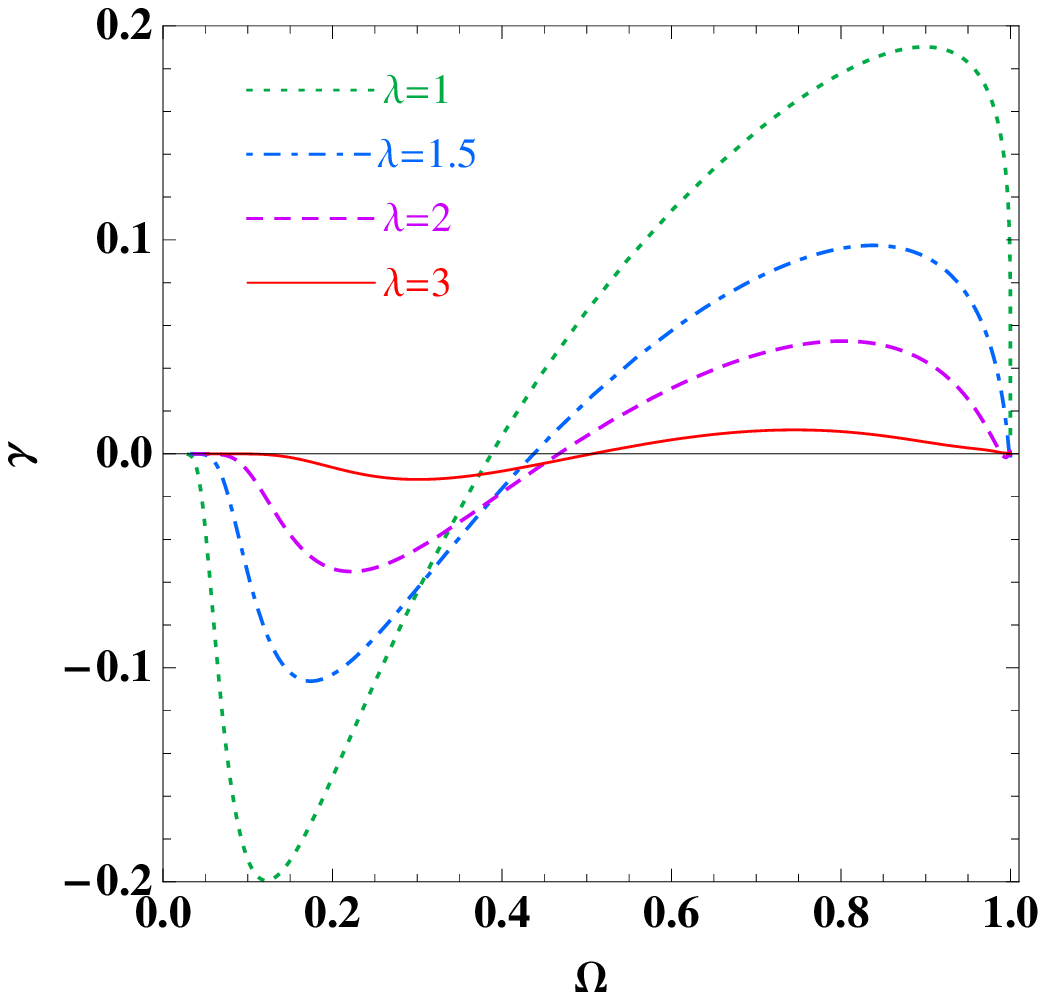} \\ [0.20cm]
\mbox{\bf (a)} & \mbox{\bf (b)}
\end{array}$
\end{center}
\caption{\small The left panel (a) shows the time evolution of model
(1) for different values of the parameters $(n,\lambda)$ in the
phase space $(\Omega,\gamma)$. For all parameters, the model is
close to $\Lambda$CDM in the past followed by a phantom phase and
the amplitude of this phase depends on the parameters of the model.
For large values of $(n,\lambda)$, the model in indistinguishable
from the $\Lambda$CDM model. The transition occurs between the
phantom and non-pantom phases around $\Omega<0.5$ (for the range of
parameters studied) before the matter-DE equivalence. The epoch of
this transition depends on the parameters of the model. For large
values of
these parameters, the transition  shifts towards the large values of $\Omega$ corresponding to small values of redshift.\\
The right panel (b) shows the time evolution of the model (2) for
various values of $\lambda$. For all models, the de Sitter point
$(\Omega,\gamma)=(1,0)$ is an attractor. } \label{fig:cosmo}
\end{figure*}

In order  to study the late time evolution in the $f(R)$ models
under consideration, we introduce the set of variable
($\Omega,\gamma,R$) ($\Omega$ is the ratio
 of the dark energy density and the critical density) and $\gamma=1+w_{DE}$

\beq
\Omega'&=&3\Omega(1-\Omega)(1-\gamma)\\
\gamma'&=&-\frac{1}{\Omega}+(3\gamma-1)(\gamma-1)-\frac{1-3\Omega(\gamma-1)}{3\Omega}\frac{R'}{R}\\
R'&=&-\frac{1}{\epsilon_{,RR}}\left[\Omega+\epsilon_{,R}+\frac{\frac{\epsilon}{R}-\epsilon_{,R}}{2}\left(1-3\Omega(\gamma-1)\right)\right],
\eeq

where a prime indicates differentiation with respect to $\ln~a$. One can see this an autonomous system of equations involving the observable cosmological parameters like $\gamma$ and $\Omega$.  \\

In what follows, we shall be interested in the f(R) models which
contain a standard matter phase \cite{Amendola:2006we} to be
compatible with the early universe physics (BBN)
($\epsilon(R)\rightarrow {\rm Constant}$ in the past) and give rise
the de-Sitter attractor at late times. Consistency also demands that
these models should satisfy the stability criteria
($(1+\epsilon_{,R}>0)$; ($\epsilon''(R)>0$\cite{Dolgov:2003px}) and
be consistent with the local gravity constraints \cite{LGC}.

Bearing in mind the aforesaid, we consider the following models:
\begin{itemize}
\item (A)  $\epsilon(x)=-\lambda R_c ~x^{2 n}/(x^{2n}+1)$ \cite{Hu:2007nk},
\item (B)  $\epsilon(x)=-\lambda R_c ~\left(1-(1+x^2)^{-n}\right)$ \cite{Starobinsky:2007hu},
\item (C)  $\epsilon(x)=-\lambda R_c ~\left(1-e^{-x}\right)$ \cite{Linder:2009jz},
\item (D)  $\epsilon(x)=-\lambda R_c ~{\rm tanh}(x)$
\cite{Tsujikawa:2007xu}'
\end{itemize}

where $x=R/R_c$ and $R_c$ of the order of the observed cosmological
constant.

Investigation reveals that
 models  (A), (B) and  (C), (D) are
 cosmologically duplicate. Thus in the analysis to follow, we shall
 focus on models (A), (C) and would refer to them as model (1) and  model (2) respectively.

\section{Cosmological Evolution}

We have investigated models (1) $\&$ (2) numerically. To do this, we have assumed that initially the models are closed to $\Lambda$CDM with $\gamma_{i} = 0$ and $\Omega$ is negligibly small. We find that
for all viable models belonging to these two classes, we have the
same evolution in the $(\gamma,\Omega)$ plane. Fig.(\ref{fig:cosmo})
shows that for both models, we have a phantom phase in the past
(small $\Omega\equiv \Omega_{\rm DE}$) giving rise to violation of
weak energy condition. As shown in  Fig.(\ref{fig:cosmo}), both the
models suffer  a transition across the phantom divide in the past
($\Omega<0.7$). For the models (1), with smaller values of $n$, the models deviate from $\Lambda$CDM whereas for models (2), the same is true for parameter $\lambda$.

\begin{figure*} \centering
\begin{center}
$\begin{array}{c@{\hspace{0.4in}}c}
\multicolumn{1}{l}{\mbox{}} &
        \multicolumn{1}{l}{\mbox{}} \\ [0.0cm]
\epsfxsize=3.2in
\epsffile{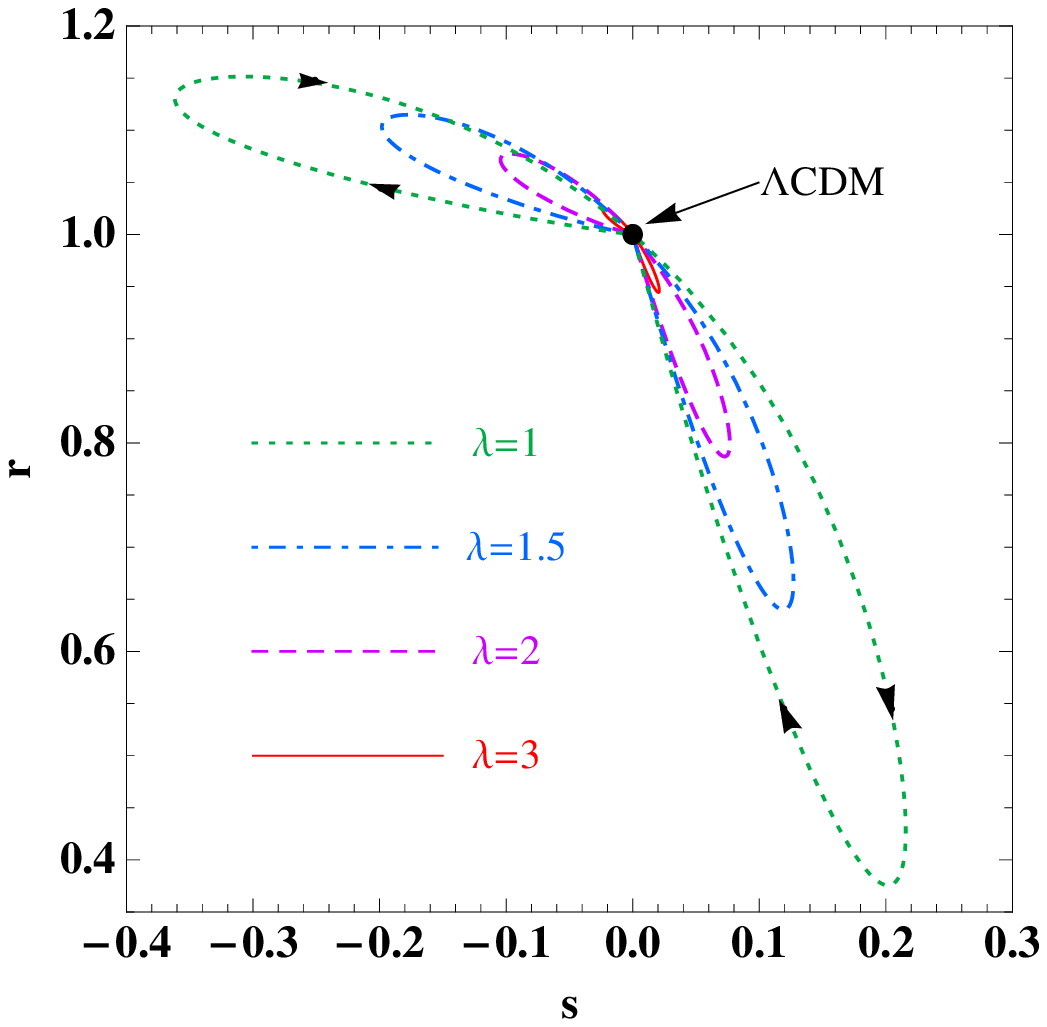} &
        \epsfxsize=3.2in
        \epsffile{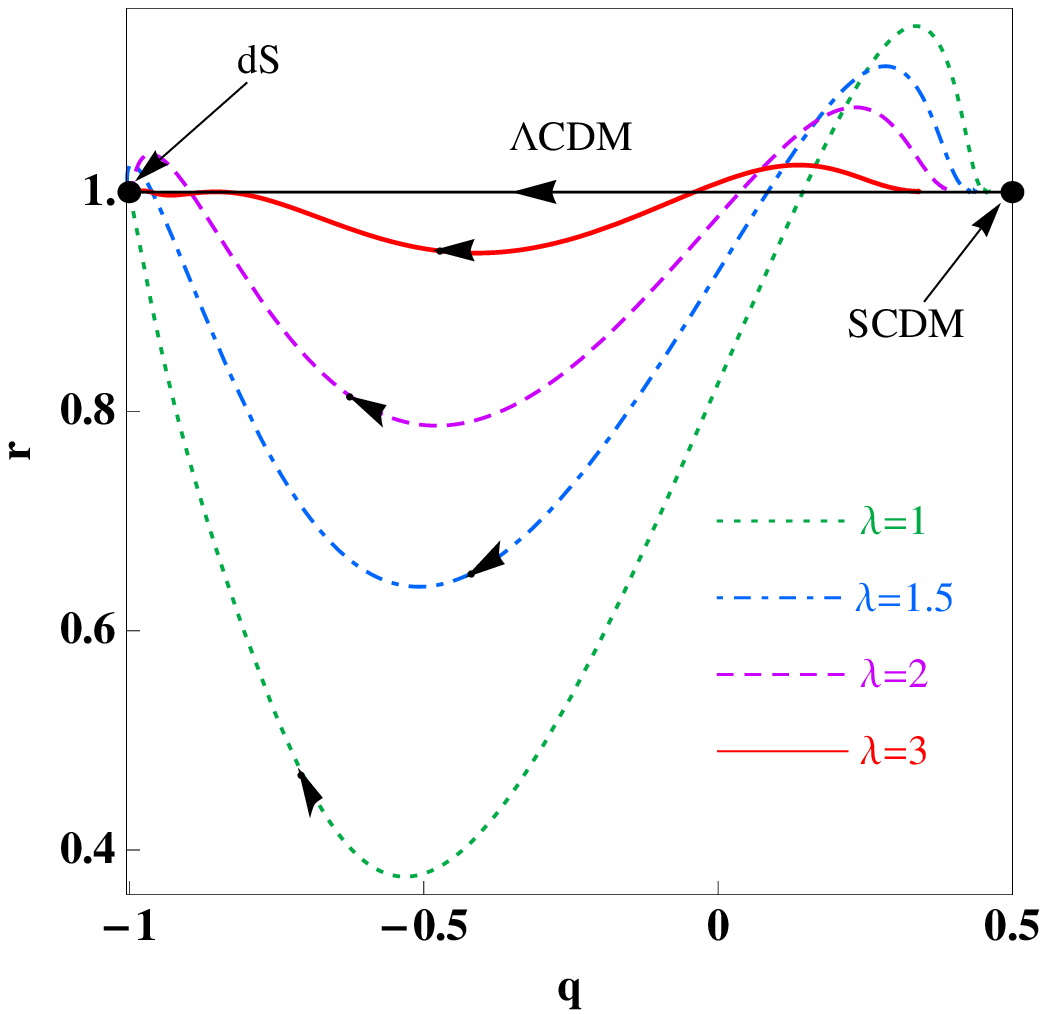} \\ [0.20cm]
\mbox{\bf (a)} & \mbox{\bf (b)}
\end{array}$
\end{center}
\caption{\small The left panel (a) shows the time evolution of the statefinder pair $\{s, r\}$ for the model (2). We have
the same evolution for various values of the free parameter $\lambda$. All models outline loops around the $\Lambda$CDM model.
The upper left part of this plane corresponds to the past evolution of the system where the model shows a phantom phase
and the lower right part is the non-pantom phase.\\
The right panel (b) shows the time evolution of the pair $\{q, r\}$.
The solid line is the time evolution of the $\Lambda$CDM model which
divides the surface into 2 planes. The upper part is the phantom
evolution of the model while the lower half is the non-pantom phase
of the model. For all models the de-Sitter (dS) point, $(s,r)=(0,1)$
or equivalently $(q,r)=(-1,1)$, is an attractor. }
\label{fig:Statefinder}
\end{figure*}

\section{Statefinder analysis}

As demonstrated in \cite{Statefinder}, it is possible to
discriminate different models of dark energy from each other using
the statefinder parameters $(r,s)$\cite{Sahni:2002fz},
 \be
r=\frac{\dddot{a}}{aH^3}=3+\frac{1-3\Omega(\gamma-1)}{2}\left(\frac{R'}{R}-1\right)
\ee where $r$ is the jerk parameter and $s$ is a function of the
jerk and the decelerating parameter $(q)$

\be
s=\frac{r-1}{3(q-1/2)}
\ee


The statefinder parameters are a natural next step beyond the Hubble function $H=\dot{a}/a$.
 By adding more derivative of the scale factor, Sahni {\it et al.}
 constructed geometrical parameters (defined using the metric only)
 which can discriminate various models. This method can be used
 to distinguish our models from  $\Lambda$CDM model, characterized by $(r,s)=(1,0)$.

\begin{figure*} \centering
\begin{center}
$\begin{array}{c@{\hspace{0.4in}}c}
\multicolumn{1}{l}{\mbox{}} &
        \multicolumn{1}{l}{\mbox{}} \\ [0.0cm]
\epsfxsize=3.2in
\epsffile{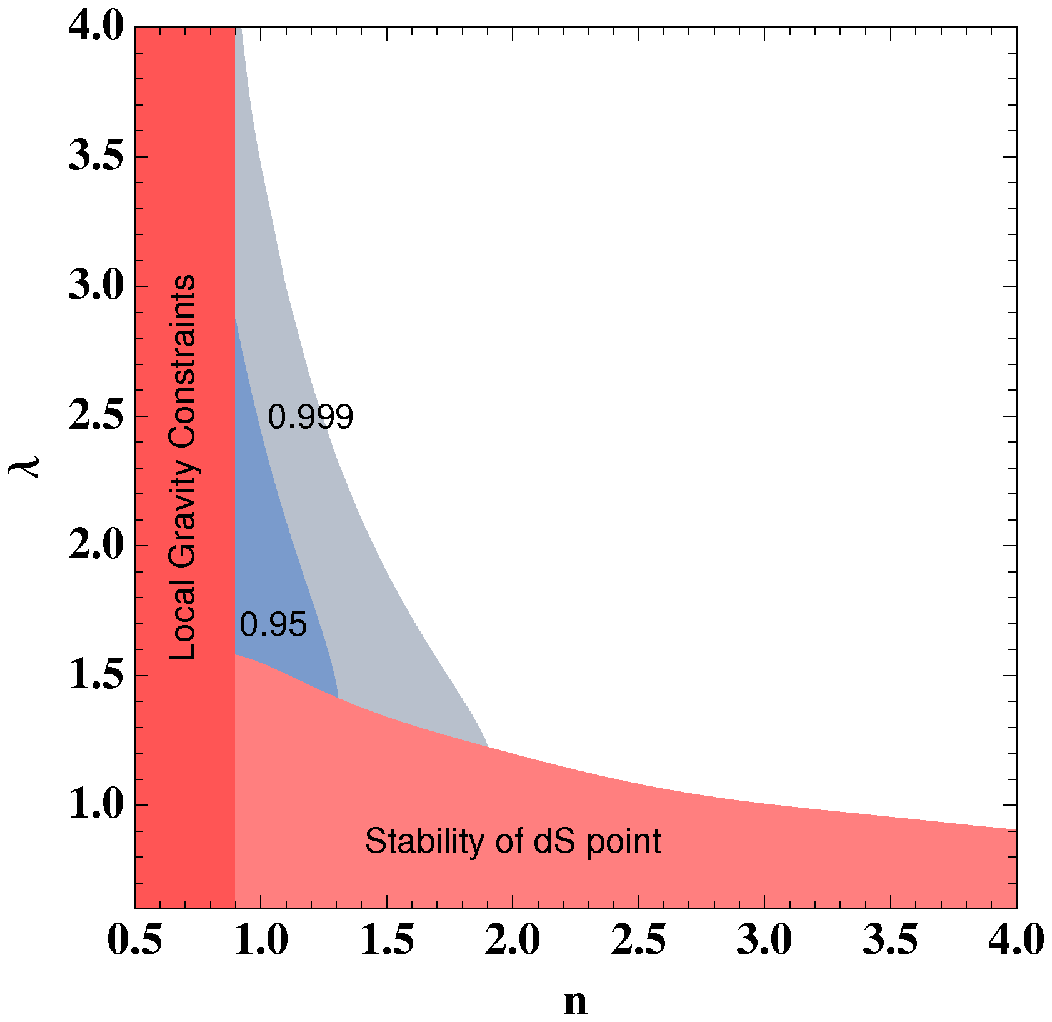} &
        \epsfxsize=3.2in
        \epsffile{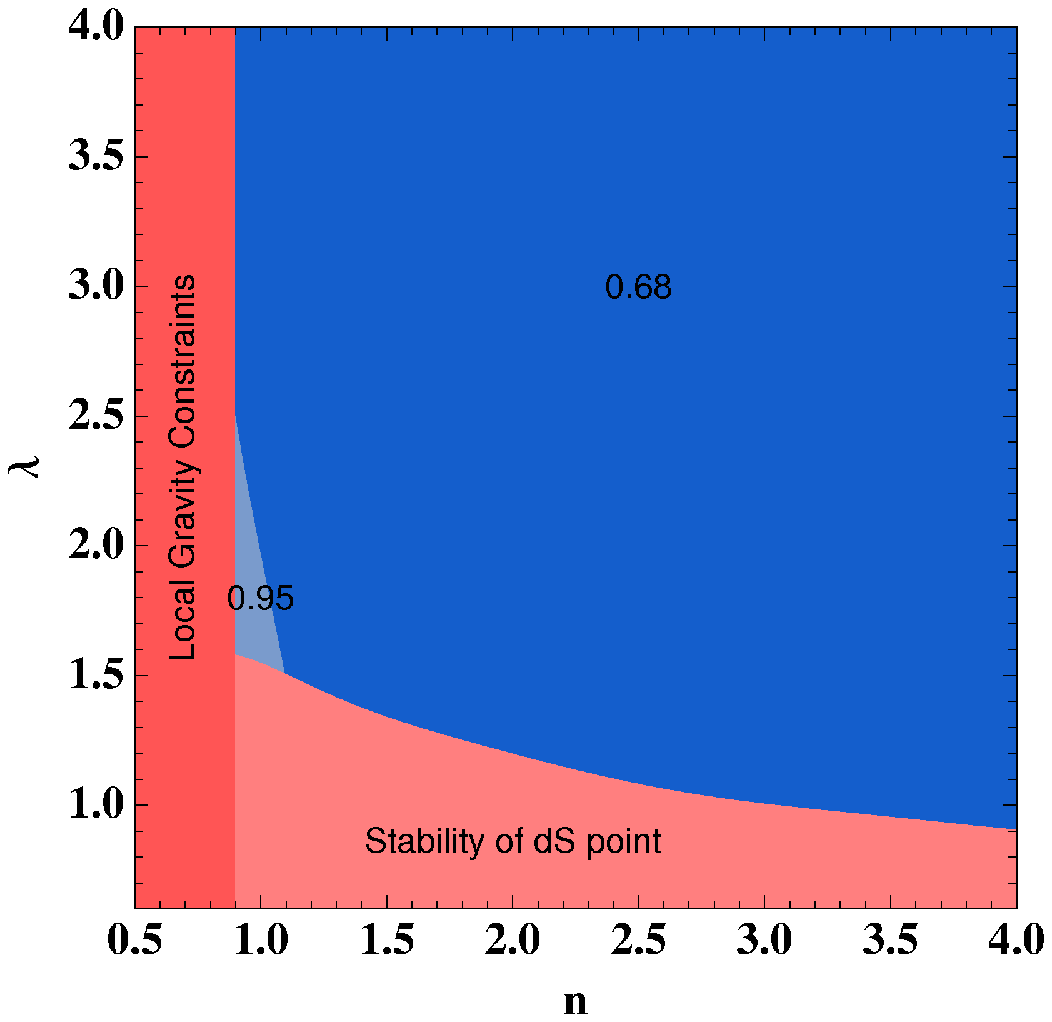} \\ [0.20cm]
\mbox{\bf (a)} & \mbox{\bf (b)}
\end{array}$
\end{center}
\caption{\small $68\%$, $95\%$ and $99.9\%$ confidence intervals for the model (1). In the left panel (a)
we imposed $\Omega_{m,0}=0.2$ and $\Omega_{m,0}=0.3$ for the right panel (b).
}
\label{fig:HS_data}
\end{figure*}

\begin{figure}
\includegraphics[height=3.0in,width=3.3in]{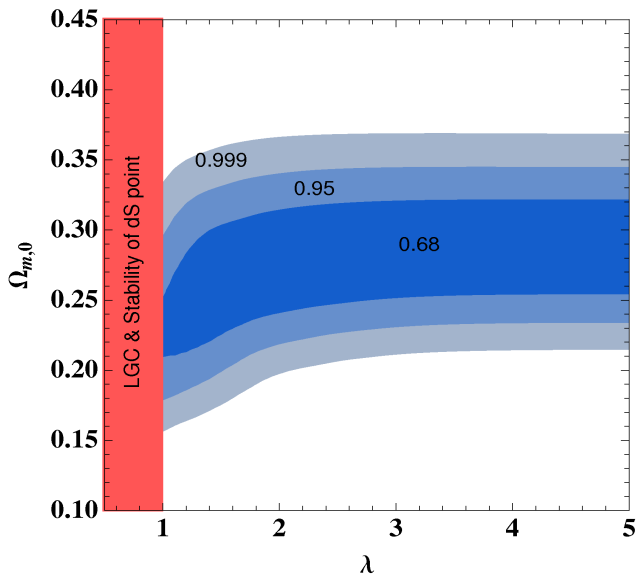}
\caption{$68\%$, $95\%$ and $99.9\%$ confidence intervals for the model (2). The local gravity constraints and the stability of the de-Sitter phase impose $\lambda>1$.
}
\label{fig:Exp_data}
\end{figure}
For both types of models (1) and (2), we found the same evolution in
the $(s,r)$ and $(q,r)$ planes. The models under consideration are
close to the $\Lambda$CDM model in the past (see
Fig.\ref{fig:Statefinder}) which appears clearly in the $(r,s)$ plane
and shows that the system is close to the critical point
$(r=1,s=0)$. The violation of the weak energy condition,
 which corresponds to the early evolution of the system, defined a loop in the upper left of the $(r,s)$-plane.
 In fact for a simple model like a power law evolution of the scale factor $a(t)\simeq t^{2/3\gamma}$,
 we have $r=(1-3\gamma)(1-3\gamma/2)$ and $s=\gamma$. Then a violation of the weak energy condition
 $(\gamma<0)$ is characterized by $r>1$ and $s<0$. The system crosses the phantom line
 (Fig.\ref{fig:cosmo}) given by $\gamma=0$ which corresponds to the $\Lambda$CDM point in the $(r,s)$-plane.
 An another loop is exhibited by the model for the quintessence evolution of the system which
 corresponds to $r<1$ and $s>0$ in the simplest power law model.\\
It must be emphasized that the aforesaid features of cosmological
evolution  appear for all the viable models in $f(R)$ that we have
studied.  These models generically exhibit two different dynamical
regimes, a phantom evolution in the past followed by quintessence
like phase at late time.

\section{Observational constraints}
We constrain the free parameters of the models studied by using
Supernovae data and the BAO data. We used the compiled Constitution
set \cite{Hicken:2009dk} of 397 type Ia supernovae for which the
$\chi^2$ is defined by

\be
\chi^2_{SN1a}=\sum_{i}\frac{(\mu_{\rm th,i}-\mu_{\rm obs,i})^2}{\sigma_i^2}
\ee

with

\be
\label{eq:mu}
\mu_{\rm th,i}=5\log\left(d_L(z_i)\right)+\mu_0+\frac{15}{4}\log\left(\frac{G_{\rm eff}(z_i)}{G_{\rm eff}(z=0)}\right)
\ee

where $\mu_0=25+5\log\left(\frac{cH_0^{-1}}{\rm M_{pl}}\right)$ is marginalised \cite{DiPietro:2002cz,Lazkoz:2005sp} and $d_L$ is the luminosity-distance.\\
The addition of the last term in (\ref{eq:mu}) takes into account a varying gravitational constant \cite{Geff}. We will not include this term in the numerical analysis. In fact it is negligible in viable $f(R)$-gravities models because of the thin-shell effect.\\

We also used the BAO distance ratio $D_v(z=0.35)/D_v(z=0.2)=1.736\pm 0.065$ \cite{Percival:2009xn}, where

\be
D_v(z)=\left[\frac{z}{H(z)}\left(\int_0^z \frac{\rd z'}{H(z')}\right)^2\right]^{1/3}
\ee

In case of  model (1) we fixed the value of $\Omega_{m,0}$ (Fig.
\ref{fig:HS_data}) and find that the model is strongly sensitive to
this parameter. For $\Omega_{m,0}=0.2$, the small values of the
parameters $(n,\lambda)$ are preferred. This is the range of the
scalar regime \cite{Tsujikawa:2009ku} which is crucially different
from  $\Lambda$CDM model. While the model is totally unconstrained
for $\Omega_{m,0}=0.3$.
\begin{figure*} \centering
\begin{center}
$\begin{array}{c@{\hspace{0.4in}}c}
\multicolumn{1}{l}{\mbox{}} &
        \multicolumn{1}{l}{\mbox{}} \\ [0.0cm]
\epsfxsize=3.2in
\epsffile{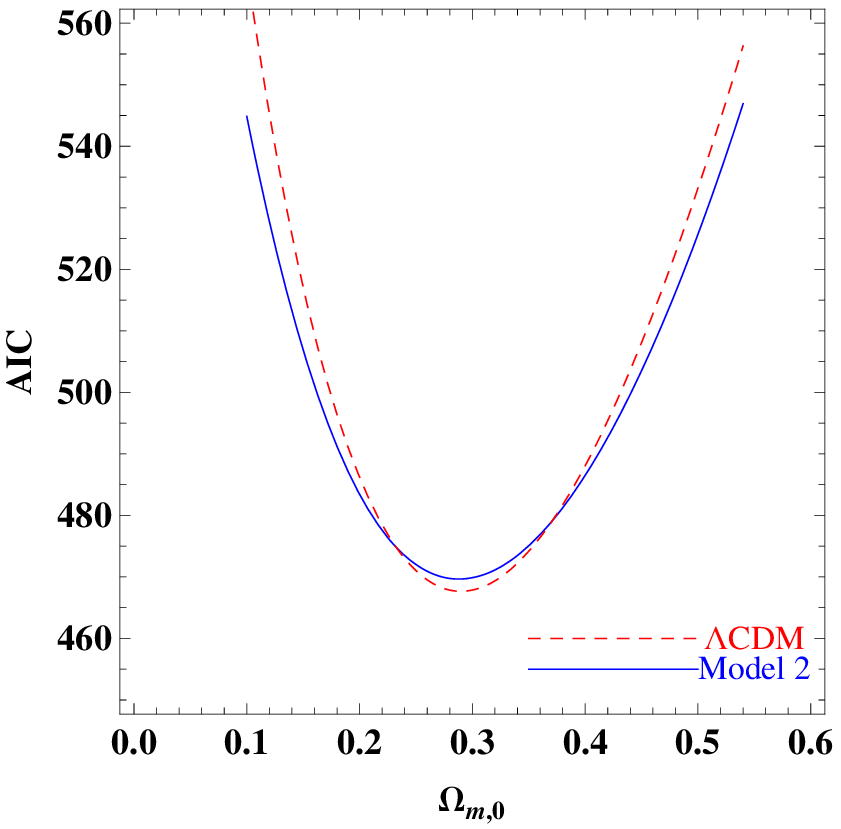} &
        \epsfxsize=3.2in
        \epsffile{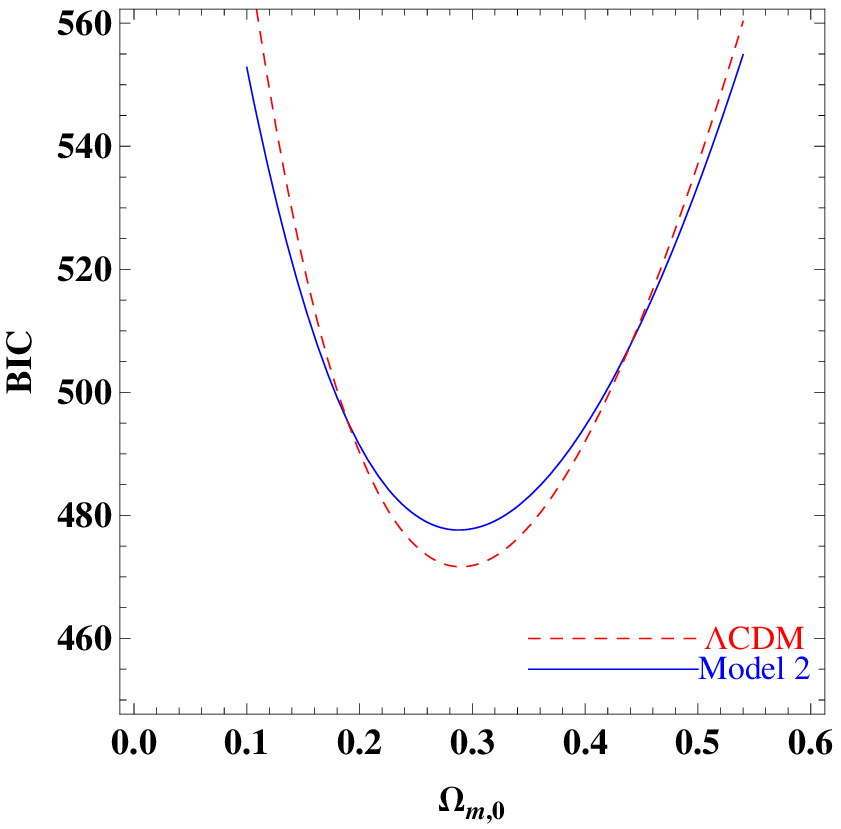} \\ [0.20cm]
\mbox{\bf (a)} & \mbox{\bf (b)}
\end{array}$
\end{center}
\caption{\small AIC(left panel)and BIC(right panel) $versus
\Omega_{m,0}$ for the standard case of cosmological constant and for
the model 2 } \label{fig:aic}
\end{figure*}

We observe that the model 2 is unconstrained by the data
(Fig.\ref{fig:Exp_data}). The density of matter today is constrained
around the concordance value and $\lambda$ appears like a free
parameter.

We also use information criteria (IC) to assess the strength of
models. These statistics favors models that give a good fit with
fewer parameters. We use the Bayesian information criterion (BIC)
and Akaike information criterion (AIC) to select the best fit
models. The AIC and BIC are defined as
\begin{equation}
AIC = -2 ln L + 2k
\end{equation}
\begin{equation}
BIC = -2 ln L +k ln N
\end{equation}
Where L is the maximum likelihood, k is the number of parameters and
N is the number of data used in the fit. For gaussian errors,
$\chi^2 = -2 ln L$, we plot the best fit values of the AIC and BIC
as the function of $\Omega_{m,0}$ for the standard model based on
cosmological constant and model 2 respectively (Fig.5) (For model (1) this has been already done in \cite{amendextra}. We can see
from the figure  that while the cosmological constant gives slightly
better fit, when larger or smaller values of $\Omega_{m,0}$ are
considered the AIC and BIC tests shows model 2 is slightly favoured
over $\Lambda$CDM. . For model 2, $\Delta AIC = 2$ and  $\Delta BIC = 6 $.

\section{Conclusion}
In this paper, we have examined the cosmological dynamics of two
different classes of viable $f(R)$-gravity models.
To begin with, we have formed an autonomous system of equations involving
observable cosmological parameters like $\gamma$ and $\Omega$ together with the
Ricci scalar $R$. This is interesting set of equations and can give interesting results upon phase-plane analysis.

We then use this system equations to study the cosmological behaviour of the models. For this we assume that in the early time the models are close to $\Lambda$CDM. As a generic
feature, these models exhibit two distinct regimes in the
time-evolution of the system. In the past, the models  violate the
weak energy condition {\it a la} the phantom phase followed by a
quintessence like behavior at the present epoch. A simple analysis
at small redshift can lead to wrong conclusion that the quintessence
model fits the data perfectly.
Fitting our models with SnIa data as well as with the BAO, we see that except for small $\Omega_{m}$ where there is strong bound on $n$ for models (1), there is no significant constraints on both the models from cosmological observations.

We emphasize that a comprehensive
analysis should  be conducted  independently in two different ranges
of the redshift, i.e., after and before the equivalence between
dark-energy and matter. In our opinion, this could be a smoking gun
for models which with different evolutionionary phases of dark
energy.

\section{Acknowledgements}
RG thanks the Centre for Theoretical Physics, Jamia Millia Islamia, New Delhi for hospitality.AAS acknowledges the financial support provided by the University
Grants Commission, Govt. Of India, through the major research project
grant (Grant No:  33-28/2007(SR)).


\end{document}